\documentstyle[12pt]{article}
\begin{document}

\begin{titlepage}
\begin{center}
%{\Large {\bf Universidad Nacional Aut\'onoma de Mexico\\
%Instituto de F\'{\i}sica, Departamento de F\'{\i}sica
%Te\'orica}}\\
%\vspace*{1cm}
\hspace*{10cm} Preprint IFUNAM FT-93-019\\
\hspace*{10cm} June 1993\\
%\hspace*{10cm} (Russian version: December 1992)
\vspace*{15mm}

{\Large{\bf Vector Particle Interactions\\
In the Quasipotential Approach}}\\
%$\,^{\star}$}}\\

\vspace*{10mm}

{\large{\bf Valeri V. Dvoeglazov}}$^{*,\,\dagger}$\\
\vskip1mm
{\it Depto de F\'{\i}s.Te\'orica, Instituto de F\'{\i}sica\\
   Universidad Nacional Aut\'onoma de Mexico\\
   Apartado Postal 20-364, 01000 Mexico, D.F. MEXICO}\\
\vspace*{3mm}
{\large{\bf Sergei V.  Khudyakov}}$^{*,\,\ddagger}$\\
\vskip1mm
{\it Laboratory of Theoretical Physics\\
   Joint Institute for Nuclear Research\\
   Head Post Office, P. O. Box 79, Moscow 101000 RUSSIA}\\
\end{center}
\vspace*{10mm}
\begin{abstract}

The composite system, formed by two $S=1$ particles, is
considered.
The field operators of constituents are transformed on the $(1,0)
\oplus (0,1)$ representation of the Lorentz group. The problem
of interaction of $S=1$ particle with the electromagnetic field is
also
discussed.
\end{abstract}

\vspace*{12mm}
%\centerline{{\bf Submitted to "Izvestiya VUZov: fizika (Sov. Phys. J.)"}}
\vspace*{12mm}

\vspace*{-2mm}

\noindent
----------------------------------------------------------------\\
%\vspace*{-2mm}
\footnotesize{
\noindent
$^{*}$ On leave from: {\it Dept. Theor.} \& {\it Nucl. Phys.,
Saratov State University and Sci.} \& {\it Tech. Center for
Control
and Use of Physical Fields and Radiations, Astrakhanskaya str. ,
83,\,\,
Saratov 410071 RUSSIA}\\
\noindent
$^{\dagger}$ Email: valeri@ifunam.ifisicacu.unam.mx\\
\hspace*{15mm}dvoeglazov@main1.jinr.dubna.su\\
\noindent
$^{\ddagger}$ Email: khud@theor.jinrc.dubna.su\\
\hspace*{15mm}postmaster@ccssu.saratov.su}

\end{titlepage}
\newpage

The descriptions of vector particle by  Duffin-Kemmer's
formalism,
Bargmann-Wigner's one and
canonical formalism, that is Proca's theory, are well known and
are perpetually
used. The Weinberg's $2(2S+1)$- component theory for  wave
function (WF) of
$S=1$ particles~\cite{1}-\cite{3} is not as popular as it deserves to
be. In
this approach the WF of vector bosons is written as  six
component column. It
satisfies to a single motion equation with no auxiliary conditions
needed~\cite{3}:
\begin{equation}
\left[\gamma_{\mu\nu} p_{\mu} p_{\nu}+p^2+2M^2\right]
\Psi^{(S=1)} (x) = 0,
\end{equation}
where $\gamma_{\mu\nu}$ 's are  covariantly defined
$6\otimes6$- matrices,
$\mu,\nu=1\ldots 4$.

The 6- component WF's  $\Psi^{(S=1)}={\chi \choose \phi}$ are
transformed on
the $(1,0)\oplus(0,1) $ representation of the Lorentz universal
covering group
$SL(2,C)$ .  This way of description of the $S=1$ particle has
some advantages,
indeed~\cite{3}.

Since the explicit form of $\gamma_{\mu\nu}$- matrices is known
(see e.g.~\cite{3a}), it gives the opportunity to transform
the above equation into the system of two equations for $\chi$ and
$\phi$:
\begin{eqnarray}
\cases{\left [p^2 + 2ip_{4}(\vec S\vec p) - 2(\vec S\vec p)^2\right
]\phi +
(p^2 + 2M^2)\chi= 0,& $ $ \cr
\left [p^2 - 2ip_{4}(\vec S\vec p) - 2(\vec S\vec p)^2\right ]\chi +
(p^2 +
2M^2)\phi = 0.& $ $}
\end{eqnarray}

In the case of massless particles ($M^2=-p^2=0$) the above
equations could be
rewritten in a following form:
\begin{eqnarray}
\cases{2\left[ip_{4} - (\vec S\vec p)\right](\vec S\vec p)\phi = 0,&
$ $\cr
2\left[ip_{4} + (\vec S\vec p)\right](\vec S\vec p)\chi = 0.& $ $}
\end{eqnarray}
In order to obtain Maxwell's equations for the left- and
right-circularly
polarized radiation (see e.g. the Eqs. (4.21) and (4.22) in the paper
[1b]) it
is necessary to assume\footnote{In fact, the $(1,0)\oplus (0,1)$
representation
is the bivector representation of the $SL(2,C)$ group. A bivector
can be
decomposed in the Pauli algebra as the sum of vector and
pseudovector~\cite{3b}. However the interpretation of $\Psi$
similarly to
Ref.~\cite{3b} generates the second solution of Eq. (1), what is
undesirable.}:
\begin{eqnarray}
\cases{(\vec S\vec p)\phi = \vec E - i\vec H,& $ $\cr
(\vec S\vec p)\chi = \vec E + i\vec H.& $ $}
\end{eqnarray}
We choose the representation of the spin operator for vector
particle as
follows:\\ $(S_{i})_{jk}=~-i \epsilon_{ijk}$.
Thus, we can  find out the physical meaning of  the components of
6- spinor:
\begin{eqnarray}
\cases{\phi_{k} = \tilde A_{k} - iA_{k},& $ $\cr
\chi_{k} = \tilde A_{k} + iA_{k}.& $ $}
\end{eqnarray}
In some sense they are just the combination of the  well-known
vector potential
$A_k$ and $\tilde{A}_k$, the pseudovector potential, which is
defined from the
tensor $\tilde F_{\alpha\beta} =
\frac{i}{2}\epsilon_{\alpha\beta\mu\nu}
F_{\mu\nu}$ dual to the tensor of electromagnetic
field\footnote{This potential
was introduced in Refs.~\cite{4a}.}.

We have used earlier $2(2S+1)$- formalism with an aim to find
out the covariant
three-dimensional equal-time equations for the composite systems
formed by fermion and vector boson as well as by two vector
bosons~\cite{4}.
The equations of such kind, called the quasipotential equations,
are employed
successfully  when solving both problems of scattering and
problems of bound
state theory , e.g. when calculating the energy spectra of
two-fermion bound
states~\cite{5}. The development of non-Abelian gauge theories
and their
experimental confirmation (e.g.
discovery of $W^{\pm}$ and $Z^{0}$- bosons; the data of
$e^{+}e^{-} $-
annihilation into hadrons) induce the investigations  of bound
states of these
gauge vector  bosons , e.g.  the investigation of gluonium.
Moreover, the high
spin hadronic resonances, which will become  more accessible at
CEBAF, NIKHEF,
RHIC and other new medium energy nuclear physics facilities,
require the
adequate treatment.

Now the majority of the authors ~\cite{6,7} considers an
interpretation of the
gluon as the mass particle having the dynamical mass (which is
the result of
the self-interaction of color-charged objects) to be possible.
Therefore we
suppose below that the gluonium consists of mass structural gluons
interacting
by means of  exchange by the massless  gauge gluon. The
scattering amplitude
for two interacting gluons has the following form in a second
order of
perturbation theory (we separate the Wigner's rotations):
\begin{eqnarray}\label{eq:t}
\lefteqn{T^{(2)} (\vec{p},\vec{\Delta}) = -
\frac{3g^2}{2M(\Delta_{0} - M)}
\left\{\left[ \frac{2(p_{0}(\Delta_{0} + M) + \vec
p\vec{\Delta})^2}{M^2} -
2M(\Delta_{0}+M)\right ]\times\right.}\nonumber\\
&\times&\left.\left ( A+\frac{(\vec
S_{1}\vec{\Delta})^2}{M(\Delta_{0}+M)}B
\right )\left ( A + \frac{(\vec
S_{2}\vec{\Delta})^2}{M(\Delta_{0}+M)} B \right
)+\right.\nonumber\\
&+&\left.2i\vec S_{1}[{\vec p}\times{\vec\Delta}]\left[
\frac{p_{0}(\Delta_{0}+M)+(\vec p\vec{\Delta})}{M^2}\right
]\left(
A+\frac{(\vec S_{2}\vec{\Delta})^2}{M(\Delta_{0}+M)} B\right
)+\right.\nonumber\\
&+&\left.2i\vec S_{2}[{\vec p}\times{\vec\Delta}]\left[
\frac{p_{0}(\Delta_{0}+M)+(\vec p\vec{\Delta})}{M^2}\right]
\left(
A+\frac{(\vec S_{1}\vec{\Delta})^2}{M(\Delta_{0}+M)} B
\right)+\right.\nonumber\\
&+&\left.\left[ (\vec S_{1}\vec{\Delta})(\vec S_{2}\vec{\Delta}) -
(\vec
S_{1}\vec S_{2})\vec{\Delta}^2\right] C^2
- \frac{2}{M^2} \vec S_{1}[{\vec p}\times{\vec\Delta}]\vec
S_{2}
\left [{\vec p}\times{\vec\Delta}\right ]C^2\right\},
\end{eqnarray}
The following notations are used:
\begin{eqnarray}
A&=&2 + \frac{4}{3}\frac{\Delta_{0}-M}{M} \kappa ,\\
B&=&1 - \lambda - 2\kappa - \frac{2}{3}
\frac{\Delta_{0}-M}{M}\kappa,\\
C&=&1 + \lambda + \frac{\Delta_{0}-M}{M}\kappa.
\end{eqnarray}
The quantities $\lambda$ and $\kappa$, mentioned above,
characterize the
magnetic dipole and electrical quadrupole momenta of the
particle,
respectively. $M$ is the mass of vector particle, $\vec p$ is the
initial
momentum in the c.m.s. and $\Delta_{\mu}$ is the 4- vector of
momentum transfer
in the Lobachevsky space.

In the quasipotential approach proposed by
Kadyshevsky~\cite{8a}  all particles
are taken on the mass shell (even in the intermediate states). In
such case
(also at $\lambda=0$  and $ \kappa =0$) the  substitution
\begin{equation}
\hat V^{(2)}(\vec p,\vec\Delta) = \hat T^{(2)}(\vec
p,\vec\Delta)\mid_{A=1,B=0,C=1}
\end{equation}
should be done in (\ref{eq:t}) in order to obtain the quasipotential
in a
second order of perturbation theory.
As a result  the obtained expression will coincide with the
quasipotential of
two interacting spinor ($S=1/2$) particles except for the
substitutions:
\begin{equation}
\frac{1}{2M(\Delta_{0}-M)}\rightarrow\frac{1}{\vec\Delta^{2}}
; \,\,\, \vec
S\rightarrow \vec\sigma
\end{equation}
which is a consequence of  normalization choice only. This
remarkable fact
shows out the similarity of Dirac's and Weinberg's $2(2S+1)$-
component
formalisms.

The relativistic configurational representation (RCR), which is
used for
description of  interactions of the spinor particles in~\cite{8}, is
also
applicable to the case of interacting vector particles. It is a
generalization
of $x$- representation of non-relativistic quantum mechanics. The
Shapiro
transformation~\cite{9a} should be used in order  to pass in this
representation instead of the Fourier transformation. The same
technique as in
[8a] could be applied to the problem of  two gluon bound states.
For example,
the partial-wave equation for the singlet gluonium state could be
written as
following:
\begin{equation}
\left ({\cal M}_{nl}-2\hat H_{l=J}\right )\Phi_{l=J}(r)=\left(\hat
V_{C}-2\hat
V_{S}-\frac{2}{3}J(J+1)\hat V_{LL}\right)\Phi_{l=J}(r),
\end{equation}
where $\hat H_{l}$ is the relativistic Hamiltonian for the
Shapiro's
plane-waves.\\ $\hat V_{C}$, $\hat V_{S}$, $\hat V_{LS}$ and
$\hat V_{LL}$ in
the quasipotential in the RCR,
\begin{equation}
\hat V(\vec r; p_{0},\vec p)=\hat V_C + \hat V_{LS}(\vec L\vec S)
+ \hat
V_S(\vec S_1 \vec S_2) + \hat V_T  S_{12} + \hat V_{LL}
L_{12} + \hat V_{SP}
P_{12} ,
\end{equation}
are the scalar, spin-spin and  spin-orbit terms, respectively
\footnote{The
matrix elements of the  quasipotential in the RCR were presented
by us earlier
in~\cite{9}.
We are now repeating the calculations of  the  energy spectrum of
gluonium
without any
approximations in these matrix elements. The results will be
presented
elsewhere.}.

By means of the quasiclassical quantization condition (see
{}~\cite{5}),
using information about the probable candidates for gluonium (see
Table) and
employing the various model quasipotentials,
the gluonium energy spectrum can be figure out,
what will be presented in more details in the approaching
publication.
\newpage
\begin{center}
Table. {\it The experimental data for probable candidates for
gluonium\footnote{It is taken from ~\cite{10a}.}.}\\
\vspace*{5mm}
\begin{tabular}{||l|c|l||}
\hline
\hline
 &$J^{PC}$&$n^{2S+1}L_{J}$\\
\hline
$\sigma(750)$&$0^{++}$&$1\,^{1}S_0$\\
$\imath(1460)$&$0^{-+}$&$1\,^{3}P_0$\\
$G(1590)$&$0^{++}$&$2\,^{1}S_0$\\
$\theta(1720)$&$2^{++}$&$2\,^{5}S_2$\\
$g_T(2050)$&$2^{++}$&$3\,^{5}S_2$\\
$\xi(2220)$&$2^{++}$&$2\,^{3}D_2$ or $2\,^{5}D_2$\\
$g_T(2300)$&$2^{++}$&$3\,^{5}D_2$\\
$g_T(2350)$&$2^{++}$&$3\,^{1}D_2$\\
\hline
\hline
\end{tabular}
\end{center}

As a concrete application of $2(2S+1)$- formalism we present
here the
Hamiltonian for vector particle interacting with electromagnetic
field~\cite{10}, which appears to be a generalization of
well-known Shay-Good's
Hamiltonian~\cite{2}:
\begin{eqnarray} \label{eq:ham}
\hat H&=&\left\{-\frac{e}{M}\left [(p_{\mu}A_{\mu})\left
(1+\frac{1}{3}\frac{\Delta_{0}-M}{M}\left
(1-\lambda-\frac{2}{3}\frac{\Delta_{0}-M}{M}\kappa\right
)\right
)+\right.\right.\nonumber\\
&+&\left.\left.\frac{1}{2M^2}\left (\vec\Theta \vec E\right )\left
(1+\lambda+\frac{\Delta_{0}-M}{M}\kappa \right
)-\frac{1}{2M^2}\left
(\vec{\Xi}\vec B\right )\left
(1+\lambda+\frac{\Delta_{0}-M}{M}\kappa\right )+
\right.\right.\nonumber\\
&+&\left.\left.\frac{(p_{\mu}A_{\mu})}{2M(\Delta_{0}+M)}Q_
{ik}\Delta_{i}\Delta_{k}\left
(1-\lambda-2\kappa-\frac{2}{3}\frac{\Delta_{0}-M}{M}
\kappa\right )\right ]+\right.\nonumber\\
&+&\left. 2e^{2}\left[ A^{2}_{\mu}\left (
1+\frac{2}{3}\frac{\Delta_{0}-M}{M}\kappa\right
)+A^{2}_{\mu}\frac{(\vec
S\vec\Delta)^{2}}{M(\Delta_{0}+M)}\left
(1+\frac{2}{3}\frac{\Delta_{0}-M}{M}\kappa-\frac{\kappa}{2}
\frac{\Delta_{0}+M}{M}\right )-\right.\right.\nonumber\\
&-&\left.\left.\frac{1}{M^3}(p_{\mu}A_{\mu})(W_{\nu}A_{\nu
})(\vec
S\vec\Delta)-\frac{1}{M^2}(W_{\mu}A_{\mu})(W_{\nu}
A_{\nu})\left
(1+\frac{\Delta_{0}-M}{M}\kappa\right)-\right.\right.\nonumber\\
&-&\left.\left.\frac{1}{M^2}(W_{\mu}A_{\mu})(W_{\nu}A_{\nu})
\frac{(\vec
S\vec\Delta)^2}{M(\Delta_{0}+M)}(1-2\kappa)\right ]\right
\}\otimes
D^{(1)}\left\{V^{-1}(\Lambda_{\vec p},\vec k)\right\}
\end{eqnarray}
In the Eq. (\ref{eq:ham}) the vectors
\begin{eqnarray}
\vec\Theta &=&(\Sigma_{(41)},\Sigma_{(42)},\Sigma_{(43)}),\\
\vec\Xi &=&
i\left(\Sigma_{(23)},\Sigma_{(31)},\Sigma_{(12)}\right),
\end{eqnarray}
constructed from the tensor components $\Sigma_{(\mu\nu)}(\vec
p) $,
\begin{equation}
\Sigma_{(\mu\nu)}(\vec p) = \frac{1}{2}\left \{ W_{\mu}(\vec
p)W_{\nu}(\vec p)
-
W_{\nu}(\vec p)W_{\mu}(\vec p)\right\},
\end{equation}
have been used. Here $W_{\mu}(\vec p)$ is the Pauli-Lyuban'sky
4- vector of
relativistic spin,  $Q_{ik} $ is the quadrupole momentum tensor
for vector
particle. The vector $\frac{e}{2M^3}\vec\Xi$ could be  considered
as a vector
of the magnetic momentum  for $S=1$ particle  moving with the
linear momentum
$\vec p$.

The importance of investigations presented is proved  by the fact
that although
the different ways of description of free particles are certainly
equivalent,
but it is not clear {\it a priori} that  all formalisms will give the
same
predictions at the presence of electromagnetic field. As a matter of
fact, they
give the  different predictions~\cite{2,11,12}.

Finally, the Weinberg's $2(2S+1)$- formalism, which is used in
this paper, is
very similar to the standard Dirac's approach to spinor particles
and,
therefore, seems to be convenient for practical calculations.
Unfortunately, in
this form it can be used in the second order of perturbation theory
only, since
the theory, based on the Eq. (1), has no
renormalizability~\cite{3}. In our
opinion there is still some additional modification of this
attractive theory
which, perhaps, is connected with the introduction of Higgs
particle. Probably,
the spontaneous symmetry breaking will give the opportunity to
remove the above
mentioned shortcoming similarly to the electroweak gauge theory.

The authors express their gratitude to Profs.  V G Kadyshevsky, N
B Skachkov,
Yu F Smirnov, S A Smolyansky, Yu N Tyukhtyaev, S I Bastrukov
and A P  Topchii for extremely fruitful discussions.  We  greatly
appreciate
the technical assistance of A S Rodin.
One of the authors (V. D.)  appreciates very highly excellent
working
conditions when  being staying at the Instituto de F\'{\i}sica,
UNAM. This work
has been financially supported by the CONACYT (Mexico) under
the contract No.
920193.

%\vspace*{5mm}

\end{document}